\documentclass[preprint,aps2,cite]{revtex4}

\usepackage{graphicx}

\begin{document}

\title{\,
\vspace*{4 mm}
How Large is the Intrinsic Flux Noise

of a Magnetic Flux Quantum,

of Half a Flux Quantum
and of a Vortex-Free Superconductor?
\vspace{2cm}}

\author{{}
J.~Mannhart$^1$, T.~Kopp$^1$ and Y.\,S.~Barash$^2$\vspace{0,5cm}}

\affiliation{$^1$Center for Electronic Correlations and Magnetism,
EP6, Universit\"{a}t
Augsburg, 
D-86135 Augsburg, Germany
\\ 
$^2$Institute of Solid State Physics, Russian Academy of Sciences, 
Chernogolovka, Moscow District, 142432 Russia \vspace{2cm} }
\vspace{2cm}
%\date{\today}\vspace{2cm}

\begin{abstract}
This article addresses the question whether the magnetic flux of stationary
vortices or of  half flux quanta generated by frustrated superconducting rings
is noisy. It is found that the flux noise generated intrinsically by a superconductor is, in good approximation, not
enhanced by  stationary vortices. Half flux quanta generated by $\pi$-rings are characterized by considerably larger noise.\\

\vspace{4cm}
\noindent
With gratitude in honor of the $80^\textrm{th}$ birthday of Prof.\,Dr.~K.\,A.~M\"{u}ller. 

\end{abstract}
\vspace{1cm}

\maketitle

% body of paper here %
%%%%%%%%%%

Standard electronic devices are based on the manipulation of electrodynamic
quantities such as electric charge, current, and magnetic fields. It  is
obvious that under practical conditions, which require contacts to the devices
and operation at a finite temperature $T$, all these devices are subject to
noise which arises usually from several sources. Thermal noise and shot noise
are prime examples. Digital superconducting devices operate by processing
quanta or half quanta of magnetic flux.  For some devices, in particular for qubits,  their
noise needs to be vanishingly small. Exploring whether for superconducting devices a fundamental, ultimate noise
limit exists, we ask the following questions:  Consider a  magnetic flux
quantum that is hold stationary in a hole of a superconductor cooled to
$T \ll $ $T_\mathrm{c}$ (Fig.\,1).  The flux of the vortex penetrates a pick-up coil with an area
$A_\mathrm{coil}$ of a  diameter that is much larger than the 
London penetration depth $\lambda_L$.  Is the flux $\Phi$ in $A_\mathrm{coil}$  noisy?  Or is
the flux of a superconducting vortex under practical operation conditions
completely noise-free? If it is noise-free, is it possible to realize devices
that operate with flux quanta in a noise-free mode?  Of course, being quantized,
the fluxoid \cite{London} exactly equals $\Phi_0 = h$/2$e$ at all instants. But does the flux fluctuate?  \\

To tackle this question,  we call to mind that in the
superconductor the flux line generates a screening current with density $\vec{J}(\vec{r})$ circling the
hole on a path with inductance $L$. The length scale of its penetration into the superconductor is  given by
$\lambda_L$ \cite{London, Tinkham}.  Although these
Meissner currents  are supercurrents, they fluctuate with $\Delta \vec {J}(\vec{r})$. 
This noise of the screening currents is caused, in the presence of the magnetic field of the vortex, by thermal and
quantum fluctuations of the gauge invariant phase, as well as by fluctuations of the density of
the condensate coexisting with the noisy quasiparticle system.\\

Consequently, the
magnetic flux penetrating the hole is noisy, expressed by the vector potential
noise  $\Delta \vec {A}(\vec{r})$.  The noisy
screening current and the noisy magnetic field give rise to a gradient of the
phase of the superconducting order parameter that is given by:  $$\hbar \vec
{\nabla} \varphi = {{m} \over {n \, 2 e }} \, \vec{J} + 2e \vec {A},$$ where
$m$ is the Cooper pair mass and $\vec{A}$ is the vector potential
\cite{Tinkham}.  The phase  $\varphi$  and the number of Cooper pairs $n$  are
conjugate variables, and therefore $$\Delta \varphi  \, \Delta n \gtrsim 1,$$
an uncertainty relation which holds for large $n$ \--- the error being of the order of $1/n$
(\cite{Tinkham, Nagaosa}).
\\

Because the typical order of magnitude of $n$ is
$10^{23}$ and
$\Delta n / n \sim n^{-1/2} \sim 3\cdot10^{-12}$, for any loop closed
around a vortex,  at any instant, the integrated
phase gradient has to amount exactly to an integer number of $2\pi$,
the precision being of the order of $10^{-12}$.\\

Is the total flux penetrating $A_\mathrm{coil}$ noisy? We start to answer this question by disregarding current fluctuations at the boundary of $A_\mathrm{coil}$, having in mind to return to this boundary effect later.  Also we postpone discussions of effects induced by intrinsic equilibrium noise of the current taking place in the absence of the vortex. Then, the topological constraint described 
yields the requirement that despite of
the noise of the local current density
$\Delta \vec {J}(\vec{r})$,
the noise on the total current $\Delta \vec {I}$, 
and the noise of the magnetic
field
$\Delta \vec {A}(\vec{r})$, the total flux through the loop has to equal
exactly $h$/2$e$ in a
large spectral range. This refers to all loops that
comprise the vortex completely. The noise of the current density
and the magnetic field does therefore not
generate noise in the total flux through the loop.
The noise is
quenched by the topological constraint resulting from the uniqueness of the
phase.
This noise quenching occurs by the following microscopic mechanism: if, for
example, at  position $\vec{r}$ the screening current fluctuates   to exceed
its average value, the resulting induction changes the path of $\vec{J}
(\vec{r})$ such that  its inductance is lowered to the value that keeps
the induced flux constant \--- and noise-free.  This process specifically reduces the 
noise of flux quanta, noise arising from other sources is not necessarily affected. \\

The noise is quenched over a large spectral range.  At high frequencies, the
range is limited by the gap frequency $\Delta / \hbar$:  the condensate does
not respond sufficiently to faster fluctuations.  At low frequencies,  rare
events triggered by highly energetic fluctuations  occur when the flux
line approaches the boundary of $A_\mathrm{coil}$ or even crosses it. Such
events are heralded by their large magnetic flux changes.   Because the time
interval between such jumps scales exponentially with the energy barrier for
entry of flux lines into the superconductor, these events are rare and leave
correspondingly long, noise-free periods. For many applications, such jumps do
not jeopardize device performance. Should a quantum jump, the device has to
be reset and started afresh.\\

Thus we conclude that under the assumptions made, the screening current
and the phase fluctuations do not
cause flux noise for a  superconducting vortex.
Due to the macroscopic character of the superconducting flux quantum and the
topological requirement of the order parameter phase, the noise 
generated by a vortex
has a very low value, even under practical experimental conditions. Therefore, the frequently practiced encoding of data as separate quantum states presents a route to low noise-data processing; the system can quench low-energy noise at every processing step.\\ 

It has to be mentioned that a
significant difference exists between the flux noise in a weak-link free superconducting
loop and of a loop interrupted with weak-links such as Josephson junctions. In the
latter case only the sum of the magnetic flux contribution and the phase difference
across the junctions is equal to a multiple of $2\pi $. Under these
conditions, noise in the phase difference across the junctions usually controls
the flux noise of the loop. In case the junctions are in the zero-voltage state, 
the phase difference noise is caused via the first Josephson relation by critical current fluctuations.
If the junctions are in the  voltage state,  the phase difference noise is associated via the second Josephson relation 
with the voltage noise.
Therefore, as described by the effective electric circuit of the
device (including, in particular, resistive elements) several noise sources are coupled to the flux
noise. Such important effects have been thoroughly studied
\cite{Jackel,Clarke,Larkin}. They also take place in loops with odd
numbers of $\pi$-junctions, for example, in tricrystal rings \cite{Tsuei}.
Keeping the phase difference of $\pi$ across the junction leaves almost
a half value of magnetic flux quanta for the flux through the ring. Due
to noise in the junctions, the flux noise is significantly enhanced
as compared to a loop devoid of weak-links. Nevertheless, also in this case the fluxoid is of course 
exquisitely noise-free.  Half flux quanta generated by frustrated $\pi$-loops can provide 
a highly precise and stable flux bias for quantum interference devices that also can be rapidly switched (Fig.\,2) \cite{generator}, the flux generated by these loops is, however, 
not free of noise. \\

In contrast, the results for weak-link free loops seem to suggest the principal possibility
to build noise-free
superconducting devices that manipulate flux quanta. Is this possible, indeed?
Because the flux of a vortex is, to a good approximation, noise-free, this problem boils
down to the question whether under practical conditions also a vortex-free
superconductor is free of flux noise.\\

Does a vortex-free superconductor induce noise currents
in a coil with a diameter $\gg  \lambda_L$ as shown in Fig.\,3? Yes, the superconductor generates flux noise and, associated with the flux noise, also voltage noise. Supercurrents and
quasiparticle currents that are  activated by thermal or by quantum
fluctuations even in the absence of the vortex, generate fluctuating, local magnetic fields. These fields form closed loops.
Loops that are closed within $A_\mathrm{coil}$   do not alter the flux in
$A_\mathrm{coil}$. Flux noise is, however, generated by fluctuating loops
that straddle the boundary of $A_\mathrm{coil}$ (Fig.\,3). By this intrinsic process, superconductors induce  a small but finite
magnetic flux noise  at all parts of
their surfaces. Therefore, above the surface of a superconductor, 
a cloud of minute flux and voltage noise is generated. \\

This flux noise is closely related to the electromagnetic fluctuations present in condensed matter \cite{Lifshitz}. It is controlled by the temperature-dependent quasiparticle density and phase
fluctuations and thus represents a basic property of the material. 
The flux noise varies as a function of the distance to the surface of the superconductor. We are not aware that this intrinsic flux
noise has been determined for Al at, say, 100\,mK, to give an example, or that  its spectral density has ever been calculated. Nevertheless it is obvious that
the noise is smallest for fully gapped, $s$-wave superconductors. The flux noise causes a  minute, temperature dependent, attractive contribution to the force that acts between two closely spaced superconductors.  In superconducting qubits, at finite temperature this flux noise provides an ultimately small, yet intrinsic source of decoherence that cannot be overcome. \\

The effects of a vortex on the superconducting order parameter and also the magnetic field of the vortex are screened in the depth of the superconducting ring. Because the intrinsic flux noise is controlled deep inside the superconductor by current fluctuations on the contour of integration, the intrinsic flux noise and the vortex can interact only slightly. There are several possible mechanisms to cause a small interaction. Fluctuation-induced quasiparticles can, for example, interact with the screening current of the flux line, or be subject to Aharonov-Bohm type phase shifts when circling the vortex. Due to these interactions, the contribution of a stationary vortex to the flux noise is finite \--- minute, but not exactly zero.\\

In summary, we conclude that intrinsic flux noise generated by a superconductor is stronger for superconductors with gap nodes.  A stationary vortex does, in good approximation,
not produce additional noise.  Its noise is
suppressed by the topology and the macroscopic nature of the vortex. 
The flux of half vortices generated by frustrated $\pi$-rings is characterized by a 
significantly larger amount of noise. \\

The authors are grateful to A.\,J.~Leggett for helpful discussions and to A.~Herrnberger for drawing the figures. This work was
supported by the DFG (SFB 484), by the ESF (THIOX), and by the Nanoxide program of the EC.

\newpage
\clearpage

\noindent
\textbf{References}
{}

\newpage
\clearpage

\noindent
\textbf{Figure Legends}\\

\noindent

\noindent
\textbf{Fig.\,1}\\
Sketch of the device configuration. A vortex $\Phi_0$ is pinned stationary by a hole  in a superconductor and penetrates a coil (blue) with an area $A_\mathrm{coil}$. The Meissner screening currents $\vec J (\vec r)$ are sketched in red. \\

\noindent
\textbf{Fig.\,2}\\
Sketch of a device configuration in which a standard SQUID is biased by half a flux quantum $\Phi_0 / 2$ generated by a frustrated superconducting loop ($\pi$-loop), formed, e.g., by a tricrystal ring.\\

\noindent
\textbf{Fig.\,3}\\
Sketch of a sample configuration to illustrate the flux noise intrinsically generated by a superconductor. 
Phase fluctuations generate loops of magnetic flux ($A, B$) that penetrate a detector coil (blue) with an area $A_\mathrm{coil}$. Loops that straddle the boundary of $A_\mathrm{coil}$ ($B$) generate magnetic flux noise in the
detector coil, loops that are closed well within  $A_\mathrm{coil}$ ($A$) do not.\\

     {~}
      \begin{figure}     
      \vspace{3cm}
       \hspace{-4cm}
     \includegraphics[width=20cm]{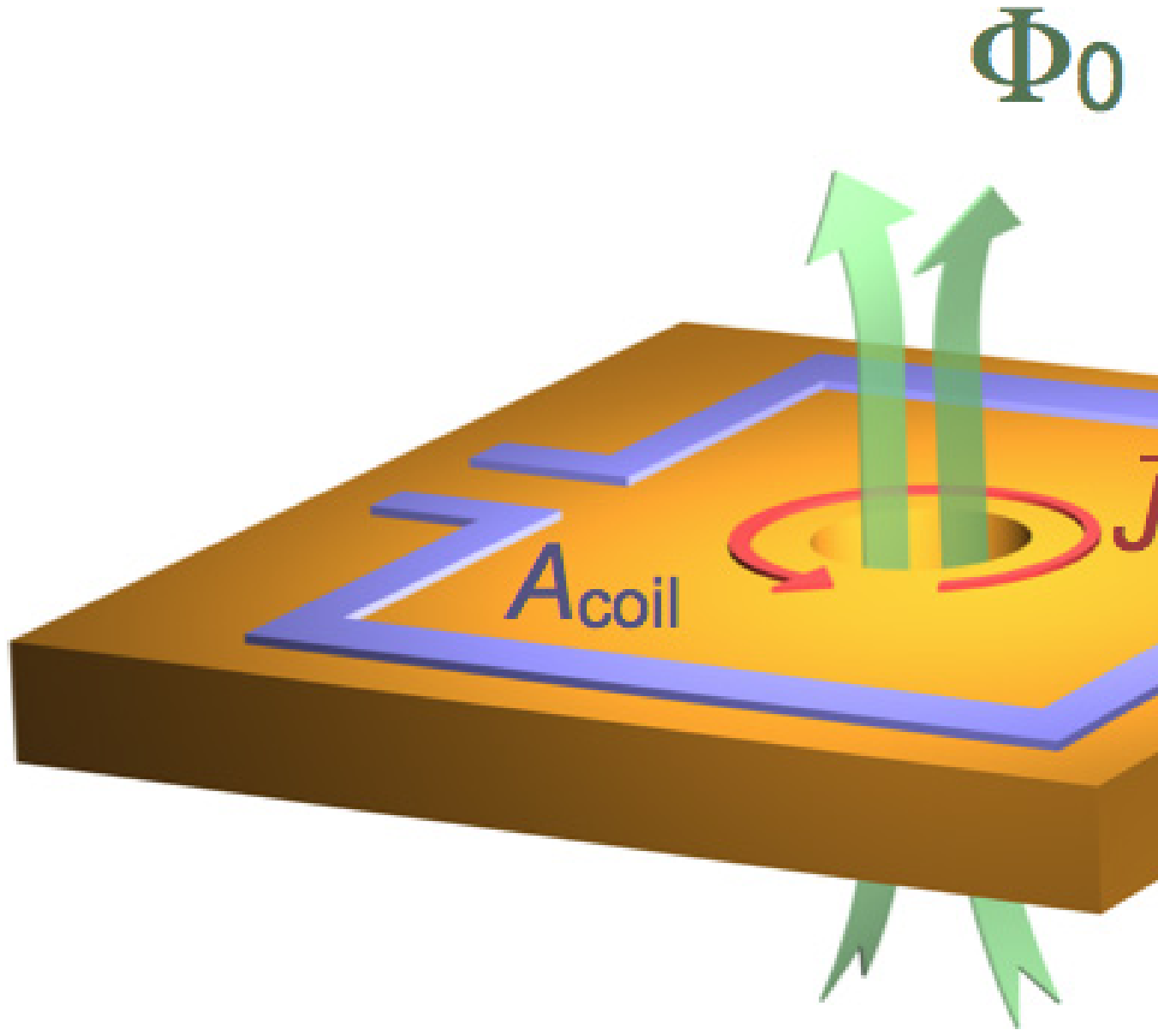}
      \end{figure}
      {~}

 {~}
      \begin{figure}
            \vspace{3cm}
      \hspace{-4cm}
     \includegraphics[width=20cm]{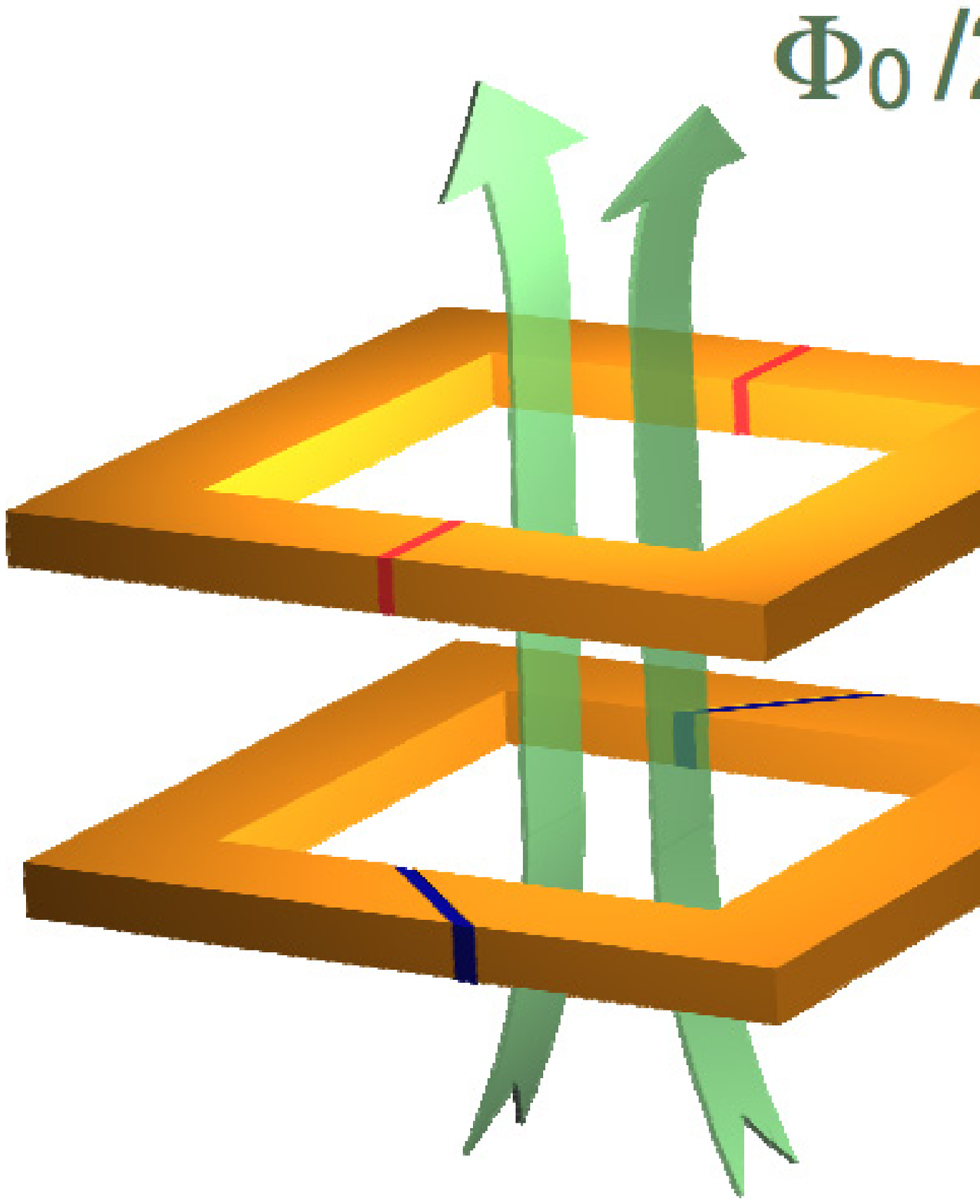}
      \end{figure}
      {~}

 {~}
      \begin{figure}
            \vspace{3cm}
      \hspace{-4cm}
     \includegraphics[width=20cm]{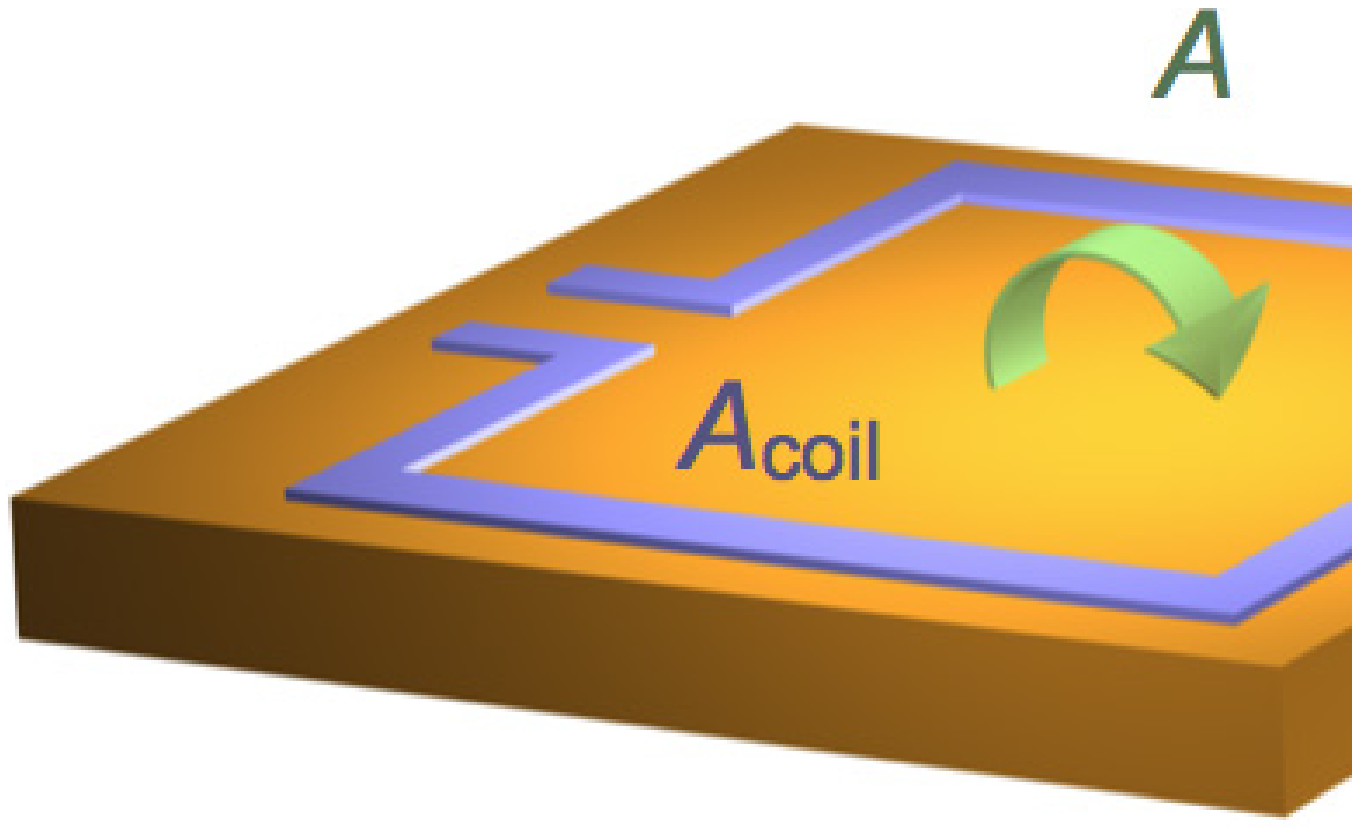}
      \end{figure}
      {~}

\end{document}